\documentclass[journal,comsoc]{IEEEtran}

\ifCLASSINFOpdf
  \usepackage[pdftex]{graphicx}
  \graphicspath{{figures/}}
  \DeclareGraphicsExtensions{.pdf,.png}
\else
  \usepackage[dvips]{graphicx}
  \graphicspath{{figures/}}
  \DeclareGraphicsExtensions{.pdf,.png}
\fi

\usepackage{booktabs}

\usepackage[dvipsnames]{xcolor}
\usepackage{comment}
\usepackage{enumitem}

\newif\ifcolormarker

\colormarkertrue

\begin{document}
\title{Graph Neural Networks for Communication Networks: Context, Use Cases and Opportunities}

\author{Jos\'e Su\'arez-Varela,
        Paul Almasan, 
        Miquel Ferriol-Galm\'es, 
        Krzysztof Rusek,
        Fabien~Geyer,
        Xiangle~Cheng, 
        Xiang~Shi, 
        Shihan~Xiao, 
        Franco~Scarselli,
        Albert~Cabellos-Aparicio, 
        and Pere~Barlet-Ros
        \\
        \vspace{0.2cm}
        \small{\textbf{NOTE:} This work has been accepted for publication in the IEEE Network Magazine.}\\
        \vspace{0.05cm}
        \scriptsize{© 2022 IEEE. Personal use of this material is permitted.  Permission from IEEE must be obtained for all other uses, in any current or future media, including reprinting/republishing this material for advertising or promotional purposes, creating new collective works, for resale or redistribution to servers or lists, or reuse of any copyrighted component of this work in other works.}
         \vspace{-1.05cm}
\thanks{J.~Su\'arez-Varela, P.~Almasan, M.~Ferriol-Galm\'es, A.~Cabellos-Aparicio, and P.~Barlet-Ros are with Barcelona Neural Networking Center, Universitat Polit\`ecnica de Catalunya, Spain.
}
\thanks{K.~Rusek is with AGH University of Science and Technology, Poland.}%
\thanks{F.~Geyer is with Technical University of Munich, Germany.}%
\thanks{X.~Cheng, X.~Shi and S.~Xiao are with Huawei Technologies, China.%
}
\thanks{F.~Scarselli is with University of Siena, Italy.}
}

%


\maketitle

\begin{abstract}
Graph neural networks (GNN) have shown outstanding applications in fields where data is essentially represented as graphs (e.g., chemistry, biology, recommendation systems). In this vein, communication networks comprise many fundamental components that are naturally represented in a graph-structured manner (e.g., topology, routing, signal interference). This position article presents GNNs as a fundamental tool for modeling, control and management of communication networks. GNNs represent a new generation of data-driven models that can accurately learn and reproduce the complex behaviors behind real-world networks. As a result, these models can be applied to a wide variety of networking use cases, such as planning, online optimization, or troubleshooting. The main advantage of GNNs over traditional neural networks lies in their unprecedented generalization capabilities when applied to other networks and configurations unseen during training. This is a critical feature for achieving practical data-driven solutions for networking. This article starts with a brief tutorial on GNNs and some potential applications to communication networks. Then, it presents two state-of-the-art GNN models respectively applied to wired and wireless networks. Lastly, it delves into the key open challenges and opportunities yet to be explored in this novel research area.
\end{abstract}

\IEEEpeerreviewmaketitle
\vspace{-0.6cm}
\section*{\large{Introduction}}
Network modeling is a fundamental component for efficient control and management of communication networks. For example, a network model enables to predict key performance indicators, such as latency, jitter, or loss, for particular network scenarios (e.g., new configuration, traffic change, network upgrade). Likewise, it can be used for autonomous network control, by pairing the model with an automatic optimization algorithm (e.g., local search, reinforcement learning)~\cite{rusek2019unveiling, vesselinova2020learning}.

Traditionally, network models have been implemented with analytical approaches, mainly based on fluid models or queueing theory. However, these models offer limited capabilities to reproduce the behavior of real-world networks (e.g., real traffic, multi-hop routing)~\cite{xu2018experience}. At the same time, accurate alternatives based on discrete-event simulation (e.g., ns-3, OMNeT++) do not scale well to large network environments. Their high computational cost limits the possibility to simulate real-world networks at scale, as well as to operate at relatively short timescales (e.g., for online optimization).

\renewcommand{\arraystretch}{1.4}
\begin{table*}[!t]
\centering
\caption{Comparison of some well-known NN models and their main applications.}
\label{table:nn-types-2}

\resizebox{1.0\linewidth}{!}{
\begin{tabular}{l p{0.45\textwidth} p{0.45\textwidth}}
\toprule
 Type & Description  & Applications \\
 \midrule
 \midrule
 Fully-connected NN & These models are \textbf{agnostic to the data type}. In fully-connected NNs, all neurons in one layer are connected to all neurons in the next layer (\textbf{all-to-all connections}). They are suitable for problems where no assumptions can be made about the input data structure.
 & \textbf{Generic feature-based classifiers} (e.g., user profile modeling, medical diagnosis, signature-based malware detection). \\
 Convolutional NN & These models assume that \textbf{the input data is spatially arranged} (typically, images). This type of NNs use convolutional kernels that \textbf{exploit the spatial dimension} of the data (e.g., they are invariant to spatial translations). & \textbf{Computer vision} (e.g., face recognition, object identification, self-driving cars), \textbf{AR/VR} (e.g., video games, telepresence services).\\
 Recurrent NN & These models assume that \textbf{the input data has a sequential structure} (e.g., text, voice, time-series). Recurrent NNs are designed to selectively keep memory of previous inputs, so that they can better  \textbf{capture sequential relationships} behind the data. & \textbf{Text and voice processing} (e.g., natural language processing, language translation, voice generation), \textbf{time-series analysis} (e.g., stock price prediction).\\
 Graph NN & These models assume that \textbf{the input data is structured as graphs} (i.e., nodes and edges). Graph NNs thus implement specific mechanisms to \textbf{exploit the relational information} behind the data (e.g.,~they are equivariant to node and edge permutation). & \textbf{Chemistry} (e.g., compound generation, drug discovery), \textbf{biology} (protein folding, protein-protein interactions), \textbf{physics} (e.g., trajectory prediction, gravitational systems), \textbf{computer Science} (e.g., \underline{communication networks}, social networks, recommendation systems).\\
 \bottomrule
\end{tabular}
}
\vspace{-0.4cm}
\end{table*}

In this context, machine learning (ML) is a promising technique for achieving accurate network models with limited execution times. In particular, deep learning models have recently gained much attention, motivated by the outstanding applications they have shown in other domains (e.g., computer vision, natural language processing).

Neural networks (NN) are data-driven. This means they can directly learn from real data, without the need for introducing theoretical assumptions as those of analytical network models (e.g., fluid models, queuing theory). As a result, they expand the possibilities to accurately model networks at a high level of detail (e.g., protocols, physical effects, hardware-level impact). In addition, these models are highly parallelizable (e.g., on GPUs, TPUs), which permits to efficiently scale to large real-world networks and big data environments.

A main aspect to produce practical data-driven solutions is generalization over networks. This refers to the capability of network models to make good predictions in different samples from those seen during training (e.g., new configurations, topology changes). In this vein, popular NN models are not suited to generalize over network-related data (e.g., fully-connected NNs, convolutional NNs, autoencoders). Communication networks comprise relational information at many different levels (e.g., topology, routing, user connections, signal interference) and these NNs are not designed to capture such type of information. The natural way to represent relational information is in the form of graphs, i.e., as sets of elements that are connected according to their relationships. Indeed, the networking community has traditionally relied on graphs as a fundamental element to represent networks and solve a plethora of control and optimization problems~\cite{vesselinova2020learning}. This eventually calls for using deep learning methods that are more suitable for graph-structured relational data.

This article posits graph neural networks (GNN)~\cite{scarselli2008graph} as a key enabler for producing accurate data-driven network models with strong generalization capabilities. GNNs are a NN family tailored to operate directly on graph-structured data. Unlike more traditional NN models, GNNs exhibit unique properties to learn and exploit relational patterns between the different elements within graphs. This property is also referred to as strong relational inductive bias~\cite{battaglia2018relational}. Eventually, it extends the possibility to accurately generalize to other networks with configurations unseen during training. As such, this new breed of deep learning models has already produced some successful applications with an unprecedented level of generalization over different types of communication networks (e.g., wired/wireless networks, data centers, IoT, SDN/NFV)~\cite{papers-gnn-github}.

\section*{\large{Graph Neural Networks}}

Graphs are an essential data type to generate structured representations of many real-life elements, such as molecules in chemistry, gravitational systems in physics, proteins in biology, or user relations in social networks. In the field of communication networks, graphs are pervasively used to represent many fundamental network components, such as the topology, routing, dependencies between flows, user connections, interference, and many others. In general, graphs enable to represent the elements that compose a network scenario (e.g., devices, users, applications) and their underlying relationships in a structured manner. This is crucial for solving many networking problems~\cite{vesselinova2020learning}.

\begin{figure*}[!t]
 \centering
   \includegraphics[width=1.0\linewidth]{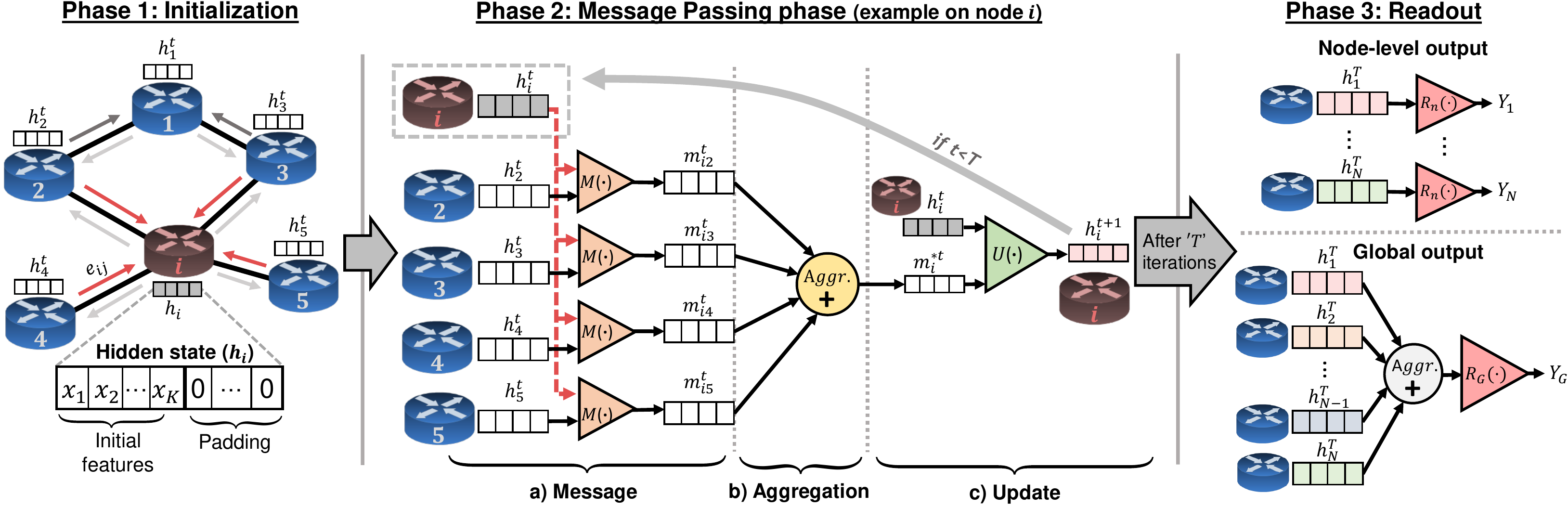}
 \caption{Schematic representation of a basic GNN model applied to a communication network.}
 \label{fig:gnn-architecture}
 \vspace{-0.2cm}
\end{figure*}

Table~\ref{table:nn-types-2} shows a comparison of some popular NN models available at the time of this writing. The most classical and generic ones are fully-connected NNs. These models are considered universal approximators that can be applied to virtually any problem and data type. Then, convolutional NNs and recurrent NNs emerged as new architectural variants to target specific data types. The main essence behind these two NN types is that they are designed to exploit relevant information from the structure of their input data. Convolutional and recurrent NNs are respectively invariant to spatial and temporal translations. These are opportunistic biases to model their target data types more accurately (i.e., images and sequences)~\cite{bronstein2017geometric}. A lesson learned from the recent history of deep learning is that generic fully-connected NNs show weaker performance when applied to problems where special-purpose NNs can be used to exploit the underlying data structure (e.g., convolutional, recurrent NNs)~\cite{bronstein2017geometric}.

GNNs~\cite{scarselli2008graph} are thus a more recent family of NN models specifically designed to process graph-structured data. These models implement a modular NN architecture that explicitly represents the elements (nodes) and connections (edges) of input graphs. Also, they introduce some inductive biases that capture meaningful patterns from the graphs seen during training (e.g., equivariance to node and edge permutation)~\cite{battaglia2018relational}. As a result, these models have already demonstrated groundbreaking applications in fields where graphs are ubiquitous (see applications in Table~\ref{table:nn-types-2}).

\vspace{-0.6cm}
\subsection*{\center{\normalfont{The GNN Architecture}}}
GNN is commonly referred to as an umbrella term for NN models that operate directly on graph-structured data. Behind the GNN concept, nowadays we can find a wide variety of NN architectural variants, which are often classified into spectral and spatial approaches~\cite{zhou2020graph}. In communication networks, the most popular GNN models are based on graph convolutional networks (GCN), graph attention networks~(GAT), or message passing neural networks (MPNN)~\cite{papers-gnn-github}. In this section, we aim to describe the architecture of a basic GNN model. To keep generality, we follow the MPNN nomenclature~\cite{gilmer2017neural}, which is considered one of the most general and expressive GNN definitions to date~\cite{zhou2020graph}. Indeed, widely used GNN models \mbox{---~including} GCN, GAT, GGNN, GraphSAGE, and many others~--- can be defined as a special case of the MPNN framework.

As shown in Figure~\ref{fig:gnn-architecture}, the execution of a MPNN can be divided into three main phases: $1)$ Initialization, $2)$ Message passing, and $3)$ Readout. We describe these phases below.

\vspace{0.2cm}
\noindent \textit{\textbf{1) Initialization} (Fig.~\ref{fig:gnn-architecture}, Phase 1):}\\
Given an input graph, the GNN generates an associated state vector for each node, known as the node's hidden state ($h_{i}$). It then initializes these states with a set of features ($X_{i}$) included in the input graph. To describe this process, we refer to an illustrative example where the input graph of the GNN is a direct representation of a wired network topology (see Fig.~\ref{fig:gnn-architecture}, left). In this example, each router represents a node in the input graph; hence the initial state features $X_{i}$ could be some router-level characteristics (e.g., switching capacity, buffer size). Hidden states $h_{i}$ are represented by n-element vectors, where the vector size is a configurable parameter of the model. Typically, these states ($h_{i}$) are larger than the initial feature vectors ($X_{i}$), so they can be simply zero-padded to fill the vector size. Likewise, in this example graph edges represent the physical links of the network, and can also be initialized with some features $e_{ij}$ (e.g., link capacity).

\vspace{0.2cm}
\noindent \textit{\textbf{2) Message passing phase} (Fig.~\ref{fig:gnn-architecture}, Phase 2):}\\
Once the routers' hidden states are initialized, an iterative process (message-aggregation-update) is executed over the input graph. Figure~\ref{fig:gnn-architecture}, Phase~2 illustrates a message passing (MP) iteration on a router (node~$i$). Note that this process would run in parallel for each router. In each MP iteration, the router combines the hidden state with its neighbors and applies three main functions along this process: message, aggregation, and update. Some of these functions are implemented by NN modules, as described later in this section.

\textit{a) \textbf{Message function} ($M$):} It encodes information about two connected nodes in the graph (i.e., adjacent routers in the example of Fig.~\ref{fig:gnn-architecture}). The $M$ function has the states of two connected routers as input (e.g., $h_{i}$ and $h_{j}$). Also, it may include some features of the link connecting them ($e_{ij}$). As a result, it produces a message ($m_{ij}$). Messages are new vectors that should encode relevant information about two connected routers and the relationship between them. For example, some properties of the traffic sent between the two routers.

\textit{b) \textbf{Aggregation function} ($Aggr$):} After messages ($m_{ij}$) are generated for all connected routers, each router combines the messages computed with its neighbors using an aggregation function ($Aggr$). This function produces an aggregated message ($m_{i}^{*}$), which is a new vector that encodes relevant information from the messages received in that node, i.e., it summarizes information from the local neighborhood. This $Aggr$ function is often implemented with an element-wise summation.

\textit{c) \textbf{Update function} ($U$):} This function updates the node states at the end of each MP iteration (Fig.~\ref{fig:gnn-architecture}, Phase~2). To do this, it combines the current state of the router ($h_{i}^{t}$) with the newly aggregated message ($m_{i}^{*t}$) --- which encodes information from the neighborhood. As a result, it produces an updated state vector for the router ($h_{i}^{t+1}$). 

Note that in each MP iteration routers only receive data from their direct neighbors. To allow routers receive information from more distant nodes in the graph, the GNN executes a number of MP iterations $T$, which is a configurable parameter of the model. For instance, in the graph of Fig.~\ref{fig:gnn-architecture}, the GNN would need at least two MP iterations to propagate information from node $1$ to node $i$.

\vspace{0.2cm}
\noindent \textit{\textbf{3) \textbf{Readout phase}} (Fig.~\ref{fig:gnn-architecture}, Phase 3):}\\
This last phase translates the information encoded in node hidden states into the final output values of the model. GNNs may typically produce two output types: $(i)$~node-level, or $(ii)$~global graph-level features. Following the example of Fig.~\ref{fig:gnn-architecture}, the GNN might predict features at the router level (e.g., buffer occupancy), or infer global network-level properties (e.g., congestion level). 
In the first case, a readout function~($R_n$) would be individually applied to each router state to produce the final outputs. In the second case, a global output could be obtained by first aggregating all router states (e.g., element-wise sum) and then applying a global readout function~($R_G$).

In general, a MPNN comprises four main building blocks, which are the four functions described earlier: message~($M$), aggregation ($Aggr$), update ($U$), and readout ($R$). Typically, the $M$, $U$, and $R$ functions are respectively approximated by three NNs (e.g., fully-connected NNs). The $Aggr$ function is often implemented with an element-wise summation. Then, GNNs are dynamically built based on the input graph. Each time the GNN receives a new graph, it combines multiple copies of the previous four functions according to the nodes and connections of the input graph. Note that, once the GNN model is assembled, it forms a recurrent network. This means it is possible to train the whole GNN model end-to-end. Hence, functions approximated by NNs ($M$, $U$, and $R$) are jointly learned across all their copies in the GNN, by applying a common backpropagation method as in any other NN model. After training, these NNs learn generic functions according to the purpose the GNN was trained for. For example, the NN that approximates the $M$ function is expected to encode relevant information in messages according to the purpose of the GNN (e.g., network anomaly detection). As a result, these generic functions can be applied to other graphs with different structures (i.e., nodes and connections) unseen during training.

\section*{\large{GNN-based Network Modeling}}

This section motivates the benefits of GNNs for building
practical data-driven solutions for network modeling.

\begin{figure}[!t]
 \centering
   \includegraphics[width=1.0\linewidth]{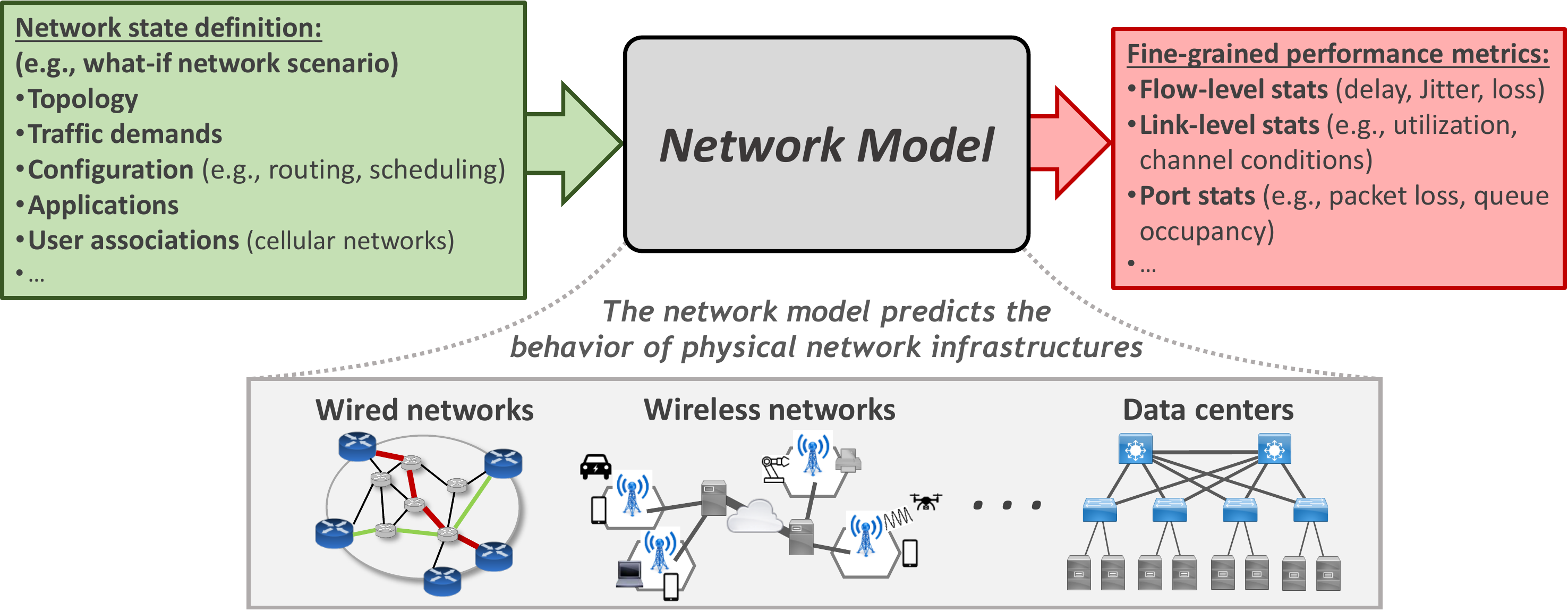}
 \caption{Schematic representation of a network model.}
 \label{fig:ndt-scheme}
\end{figure}

Figure~\ref{fig:ndt-scheme} shows a black-box representation of a generic network model. This model is given an input network scenario (e.g., topology, traffic, configuration), and it is tasked with predicting relevant performance metrics at different levels of granularity (e.g., flow, link, port statistics). In general, network models enable to predict what would be the resulting network performance under possible topology changes (e.g., upgrades, failures), or with new configurations (e.g., routing, VNF placement). These models enable a plethora of network control and management tasks, such as what-if analysis or automatic network optimization.

\subsection*{Generalization properties of GNNs over networks}

Networks comprise graph-structured information at many different levels~\cite{vesselinova2020learning}. In this vein, more traditional NN models, such as fully-connected NNs, are not designed to directly process and capture meaningful patterns from graphs. GNN thus represents the most suitable ML technique nowadays for processing such graph-structured data.

GNN models have unique properties to accurately generalize over graphs. For example, unlike other ML models, GNNs are focused on relational reasoning and combinatorial generalization over graphs~\cite{battaglia2018relational}. These models leverage the distributed message passing architecture described earlier to get local context on graph nodes. For example, in the scenario of Figure~\ref{fig:gnn-architecture} the internal NN functions of the MP phase (e.g., message, update) learn from the individual perspective of each router in the network. This feature eventually endows the model with strong generalization capabilities, as it learns from the experiences locally seen by all routers during training. As a result, it can then apply this learned knowledge to other routers in different networks with variable sizes and structures.

Moreover, GNNs are equivariant to node and edge permutations~\cite{bronstein2017geometric}. This means that, if we represent network scenarios as graphs, GNN models can find symmetries or equivalent patterns between the network scenarios seen during training and the new scenarios where the model is applied after training. Following with the example of Figure~\ref{fig:gnn-architecture}, the GNN would be able to identify clusters within the network topology that are equivalent or similar to others seen during training. In practice, different types of networks (e.g., wireless, data centers, IoT) have their own particularities and may comprise different types of elements and relationships. However, these generalization properties are extensible to any network scenario as long as it is represented as a graph.

\subsection*{Practical advantages of GNNs}

As a result of the aforementioned generalization properties, GNN unlocks fundamental practical limitations of previous ML-based solutions for network modeling:

\textbf{Offline training}: Training a deep learning model often requires big amounts of data with enough diversity to abstract deep insights during the training phase (e.g., test different configurations, inject various traffic loads). Indeed, models typically need to observe extreme cases that may break the normal operation of the network (e.g., link failures). This is because they should be able to make good estimates in case these rare events occur once deployed. For obvious reasons, it is not feasible to reproduce these extreme cases in real production networks. Hence, a more realistic approach is to train ML models offline (e.g., in a controlled testbed), and make products that are readily available for deployment in customer networks, without the need for re-training on the target network. However, this requires relying on ML models that can generalize to new networks unseen during training (e.g., new topologies, configurations, traffic patterns). This makes GNN a key enabler for achieving practical data-driven network models that can be fully trained offline.

\textbf{Testing and deployability:} Nowadays networking products are extensively tested before being deployed, as real-world networks are considered critical infrastructures. This makes it unrealistic to rely on ML-based solutions that can be trained online, as we would need strong supervision mechanisms to check the evolution of the model. From a deployability standpoint, GNNs enable to train network models offline, test their behavior under a wide range of operational network scenarios, and finally generate certifications that can determine the operational ranges where the model offers guarantees (e.g., network sizes, maximum traffic aggregates). This conforms to the standard commercialization process of networking products nowadays.

\section*{\large{Network Control and Management\\ with Graph Neural Networks}}

A GNN-based model such as the one described in the previous section has as many applications as traditional network models. Its main benefit is that it can be applied to use cases where it is relevant to produce accurate and detailed performance metrics (e.g., end-to-end delays, jitter, loss, flow completion time). This makes these models especially useful for SLA-driven network optimization tasks, where traditional modeling techniques do not meet the requirements to achieve accurate estimates with limited cost.

Figure~\ref{fig:optimization-architecture} depicts a generic optimization architecture in the context of SDN-based networks. We opportunistically use this architecture to better illustrate the operational workflow, while similar optimization mechanisms could be implemented in traditional networks with distributed control. This optimization architecture envisions two main operation modes, depending on whether there is human intervention (open loop), or not (closed loop). Based on this, two main paradigmatic applications can be differentiated:

\begin{figure}[!t]
 \centering
   \includegraphics[width=1.0\linewidth]{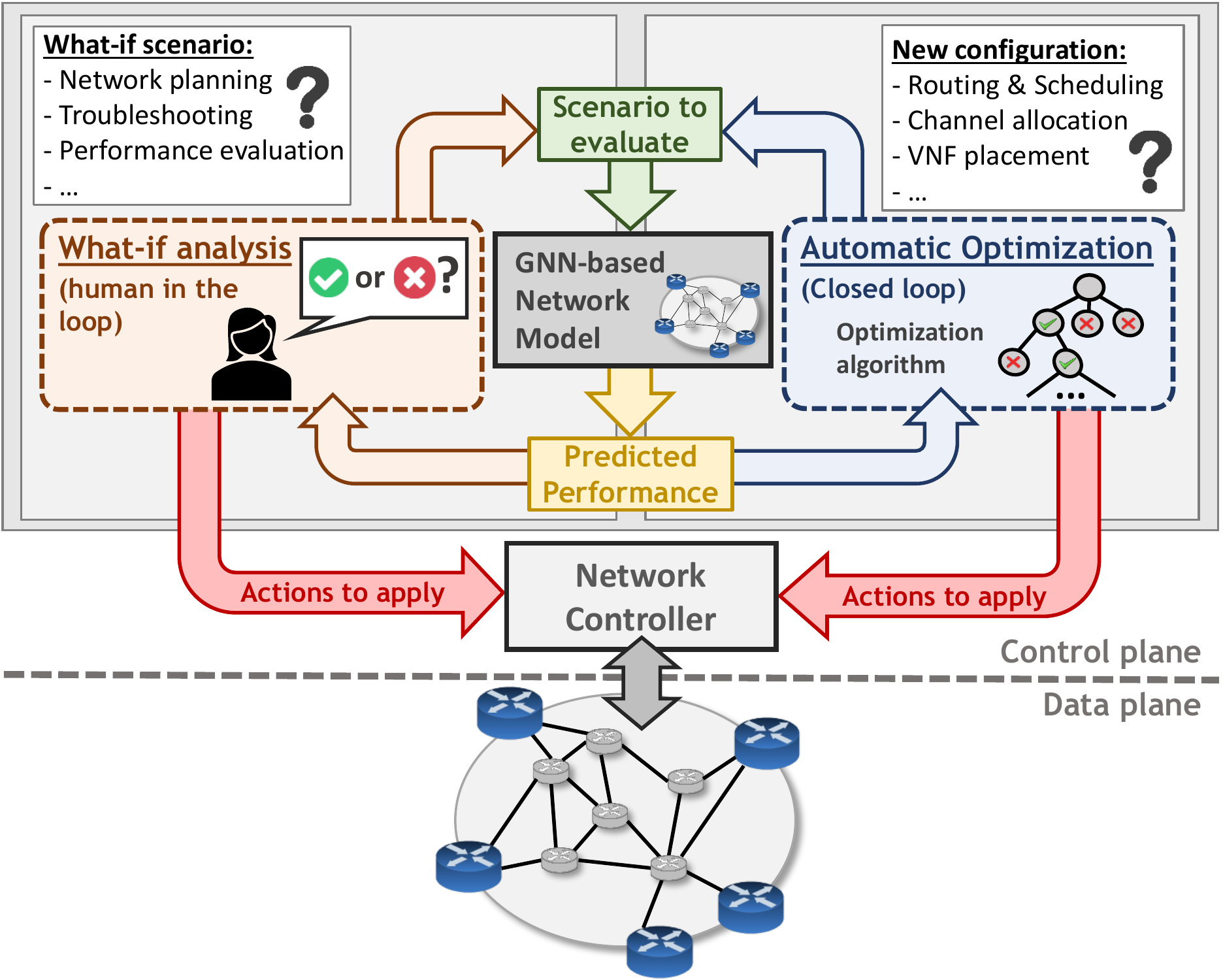}
 \caption{Optimization architecture with a GNN-based network model.}
 \label{fig:optimization-architecture}
\end{figure}

\subsection*{What-if analysis (open loop)}
This operation mode considers the case of a network administrator and/or engineer that aims to evaluate the network behavior under some specific what-if scenarios. For this purpose, the network model can be used to predict relevant performance metrics under certain scenarios (e.g., new configurations, potential failures, network upgrades). This can eventually be leveraged for a plethora of common networking tasks, such as planning, troubleshooting, optimization, or building network recommendations systems.

\subsection*{Automatic optimization (closed loop)}
This second operation mode envisions autonomous optimization tasks, with an eye on future self-driving networks. Automatic network optimization can be achieved by combining a network model with an optimization algorithm~\cite{rusek2019unveiling}. In this well-known architecture, the algorithm (e.g., local search, branch and bound, reinforcement learning) generates candidate configurations that pursue a particular optimization goal (e.g., minimize end-to-end delay). The network model is responsible for predicting the performance if these configurations were applied in the network. Thus, through an iterative generate-evaluate process the optimizer can finally produce a new configuration that meets the optimization goals.

In these types of applications, it is essential to count on a network model that can accurately estimate the target performance metrics (e.g., delay, jitter). Otherwise, there could be a problematic mismatch between the performance predicted by the model and the actual result after applying the new configuration to the network. This makes the use of analytical network models (e.g., queuing theory) arguably insufficient for control and management tasks, as these models have limited capabilities to reproduce the behavior of real networks (e.g., real traffic, physical effects, hardware-level impact). 

Likewise, from the ML perspective, the network model must generalize accurately to a broad range of network state descriptions, either produced by network administrators (what-if analysis) or by optimization algorithms (automatic optimization). In practice, this means we need a ML model that can be trained offline on a broad collection of network samples (e.g., from a controlled testbed), and that can then produce accurate estimates when deployed in new network scenarios unseen in advance (e.g., new topologies, configurations, traffic). This again calls for the use of GNNs, as they are the only ML models that offer the possibility to generalize over networks, as discussed in more detail in the previous section.

\section*{\large{Example Use-Cases}}

GNNs have already been applied to a wide variety of networking use cases, such as routing optimization~\cite{rusek2019unveiling, geyer2018learning}, Multipath TCP~\cite{zhu2020gclr}, network calculus~\cite{geyer2019deeptma}, or power control in wireless networks~\cite{shen2020graph,eisen2020optimal}. Here, we present two representative examples of custom GNN models respectively applied to wired and wireless networks: RouteNet~\cite{rusek2019unveiling}, and WCGCN~\cite{shen2020graph}. Both models are natural extensions of the canonical GNN architecture described earlier in this article. We implement these models with IGNNITION~\cite{ignnition-paper}, and make some experiments focused on showing the capabilities of these models to generalize over networks. 

\vspace{0.2cm}
\subsection*{Performance Evaluation in Wired Networks}

\begin{figure}[!t]
 \centering
   \includegraphics[width=0.71\linewidth]{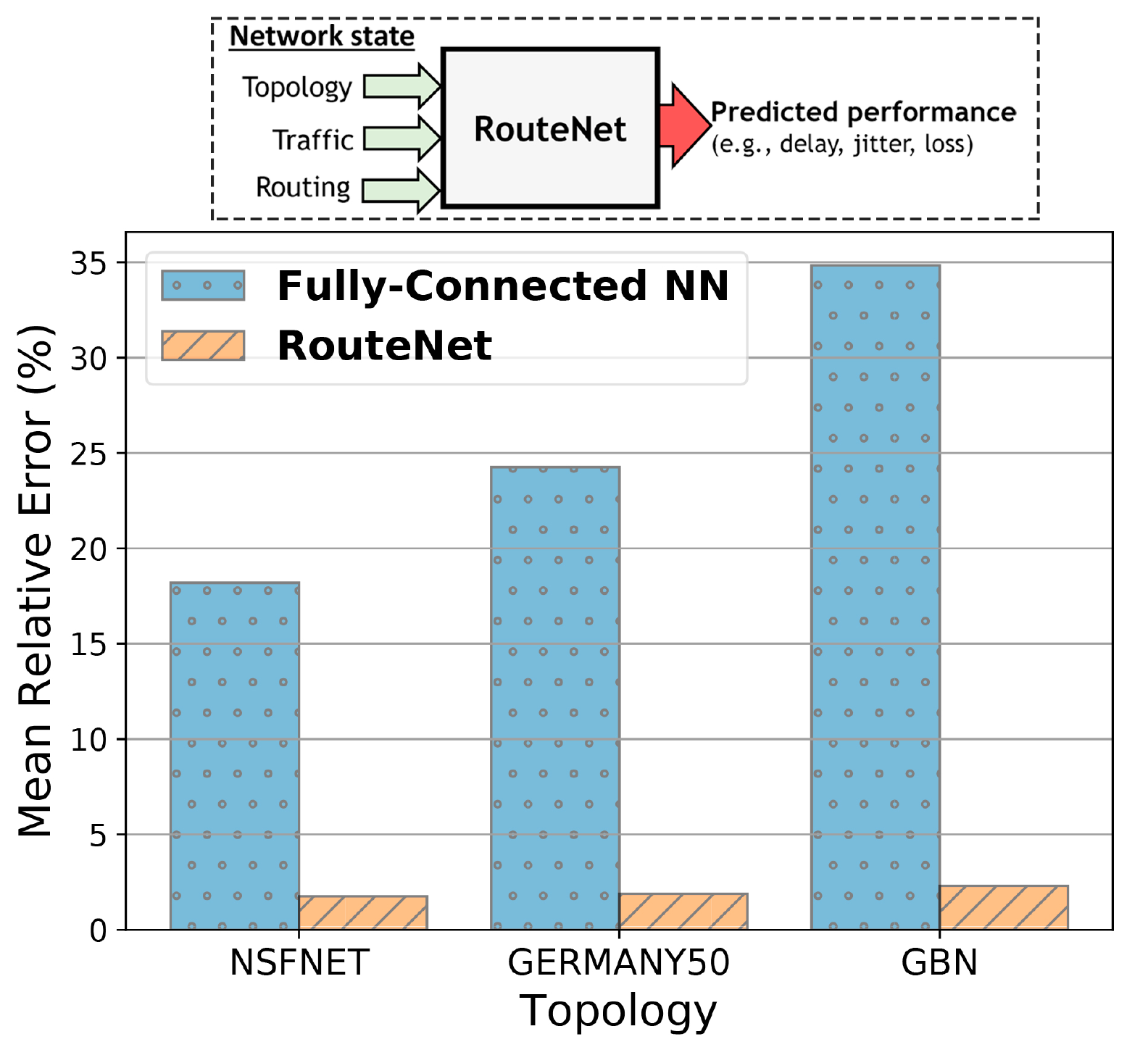}
 \caption{Mean relative error (\%) of the flow-level delay predictions made by RouteNet~\cite{rusek2019unveiling} and a fully-connected NN. Evaluated over 40,000 samples of the NSFNET (14 nodes), Germany50 (50 nodes), and GBN (17 nodes) topologies. Both models were trained only on samples from NSFNET and Germany50.}
 \label{fig:routenet}
\end{figure}

RouteNet~\cite{rusek2019unveiling} is a GNN-based model for performance evaluation in wired networks. As illustrated in Figure~\ref{fig:routenet}, this model has a network state description as input, defined by: a network topology, a traffic matrix, and a routing configuration. As a result, it produces estimates of key performance indicators at a flow-level granularity (e.g., delay, jitter, loss). This network model can be used for what-if analysis ---~e.g., to test alternative configurations~--- as well as for automatic optimization, by combining the model with an optimization algorithm (see Fig.~\ref{fig:optimization-architecture}). As an example, in~\cite{rusek2019unveiling} they leverage this model for several SLA-driven optimization use cases, such as minimizing the delay and/or jitter in the network, fast link failure recovery, or finding the optimal link upgrades.

To showcase the generalization capabilities of this model, we train RouteNet on 177,500 samples simulated in two real-world network topologies: NSFNET (14 nodes) and Germany50 (50 nodes). All these samples include a wide variety of traffic matrices and routing configurations. Then, we evaluate the predictions of this GNN model in three test datasets with 40,000 samples, respectively from NSFNET, Germany50, and a new network topology:~GBN (17 nodes).

Figure~\ref{fig:routenet} shows the mean relative error (MRE) produced by RouteNet when predicting flow-level delays on samples from the three test datasets. Here, we define the MRE with respect to the delay labels produced by an accurate packet-level network simulator (OMNeT++). Note that all these evaluation samples include combinations of routing configurations and traffic matrices unseen during training. As we can see, the model can accurately generalize to these new samples, even for those of the GBN network, which was never seen during training (MRE$<$3\%). For the sake of comparison, we repeat the same experiments with a fully-connected NN with equivalent inputs to RouteNet. In Figure~\ref{fig:routenet}, we can observe that this model produces significantly larger errors, which increase especially when applied to samples of the new GBN network unseen during training (MRE$\approx$$35\%$). These results evidence the unprecedented capability of the GNN-based model (RouteNet) to accurately generalize over all its input network parameters (i.e., topology, traffic, routing), which are internally represented in a graph-structured manner.

Beyond the accurate predictions produced by this GNN model, one main advantage is its low execution time. This may be crucial for online network optimization, as it enables testing a large set of candidate configurations in a limited amount of time. As a reference, in our experiments RouteNet takes 65 ms on average to evaluate samples of the 50-node network (Germany50). In contrast, the packet-level network simulator used to generate the datasets (OMNeT++) takes more than 10 minutes on average to simulate each sample.

\subsection*{Radio Resource Management in Wireless Networks}

WCGCN~\cite{shen2020graph} is a GNN-based model for radio resource management in wireless networks. The original paper~\cite{shen2020graph} shows how this model can be applied to maximize the sum rate on networks, which is a classical optimization problem applicable to many use cases in wireless networks (e.g., power control, beamforming).

\begin{figure}[!t]
 \centering
   \includegraphics[width=0.7\linewidth]{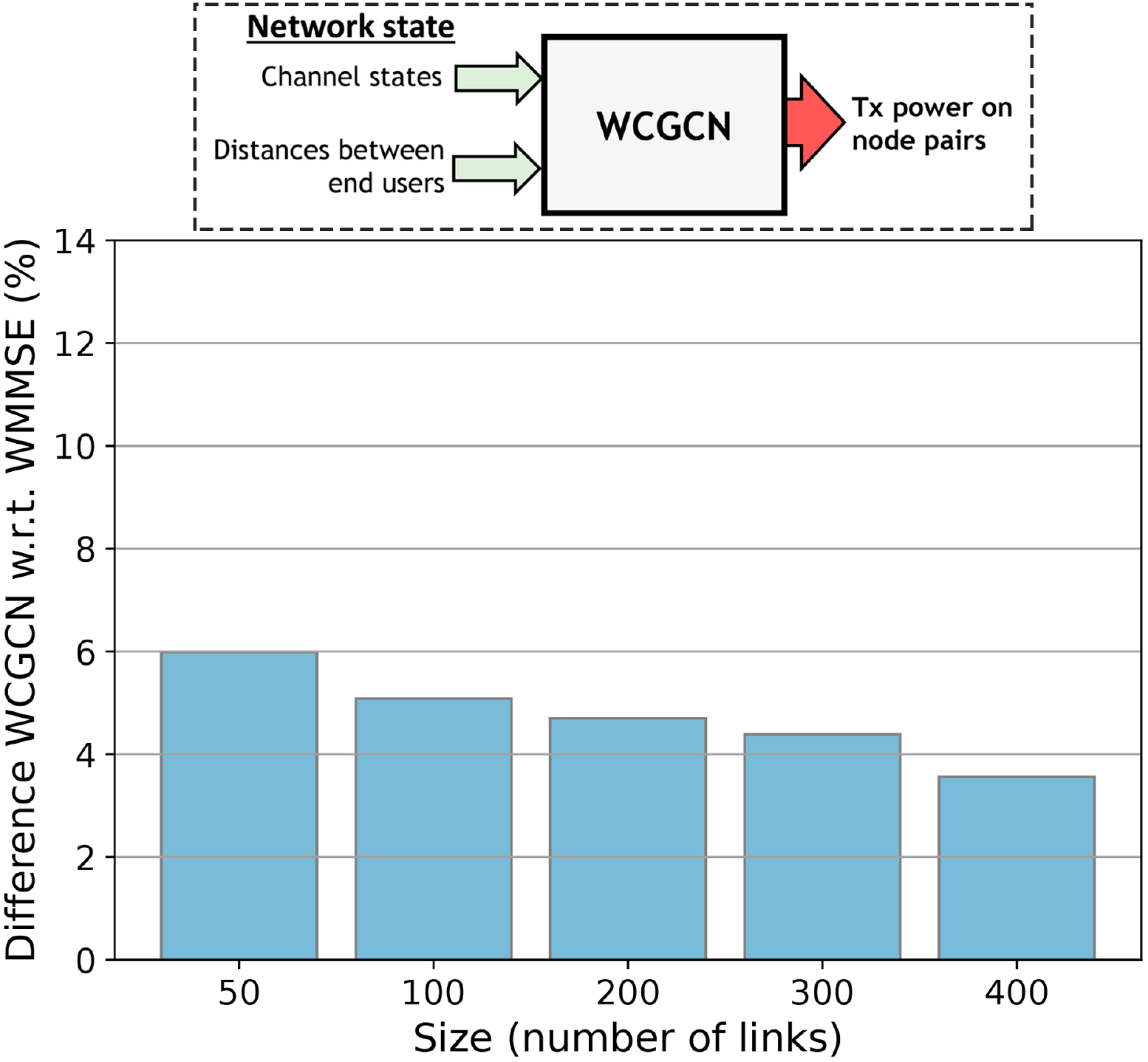}
 \caption{Mean sum rates achieved by WCGCN~\cite{shen2020graph}, normalized in \% by the performance achieved by the WMMSE optimization algorithm (100 iterations). The WCGCN model was trained on 1,000 samples with 50 links.}
 \label{fig:WCGCN}
\end{figure}

As an example, we apply this GNN model to optimize power control. The model has a description of the network state as input, which includes the channel states and the distances between end nodes. As a result, it produces the recommended Tx power on each node pair connection (see Fig.~\ref{fig:WCGCN}). We train this model on 1,000 samples of 50 links (i.e., 50 connected node pairs) and then evaluate its performance on different sets with network samples of increasing size (\mbox{50-400} links).

Figure~\ref{fig:WCGCN} shows the performance difference with respect to the classical weighted minimum mean square error (WMMSE) algorithm. This method represents a close-to-optimal approach for this problem. As we can observe, even if WCGCN was only trained on 50-link scenarios, it achieves similar performance to the near-optimal WMMSE algorithm in networks up to 400 links ($\approx$+4\%). This shows the unprecedented scalability and generalization capabilities of this ML model on the target optimization problem. Likewise, we observe a significant reduction in execution time compared to WMMSE. For example, in scenarios with 400 links, WCGCN produces results in 16 ms on average, while WMMSE takes more than 10 seconds.

\section*{\large{Opportunities and Open Challenges}}

This section outlines some opportunities and open challenges yet to be explored before achieving production-ready GNN-based solutions for communication networks.

GNNs can be virtually applied to any network-related domain, as long as it is formulated as graphs. Indeed, these models have been recently validated for different network applications, showing outstanding results in different communication paradigms (wired, wireless, SDN/NFV, IoT)~\cite{rusek2019unveiling, geyer2018learning,zhu2020gclr,geyer2019deeptma,shen2020graph,eisen2020optimal}. In this vein, there is still an ocean of opportunities to apply these novel methods to different use cases (e.g., optimization, troubleshooting, planning, what-if analysis) and expand their horizon towards other types of communications (e.g., satellite networks). This will not be a trivial task, as standard GNN models are not directly applicable to any networking use case. A thorough design process is often required to come up with a custom GNN architecture suitable for the target networking problem.

Despite the promising applications of GNNs to networking, these models are still in an early stage of technology readiness. We describe below some crucial open challenges that remain to be addressed:

\begin{itemize}[leftmargin=0.5cm]
\item \textit{\textbf{Generalization to real networks:}} To achieve production-ready solutions, it would be convenient to create GNN models that can be trained offline in network testbeds of limited size (e.g., in a networking lab), and that can then scale accurately to considerably larger real-world networks (e.g., with hundreds or thousands of nodes). Beyond the unprecedented generalization capabilities of GNNs in communication networks, current GNN models have limited capabilities to scale to large networks (e.g., due to over-smoothing in message aggregations). Indeed, scaling to large networks is a generic open challenge among ML-based solutions nowadays.

\item \textit{\textbf{Uncertainty:}} Like any other ML-based technique, GNNs are probabilistic models that entail some degree of uncertainty. Given the critical nature of network infrastructures, this can be a potential limitation for deploying data-driven solutions in real-world networks. More mature GNN-based solutions can be achieved by relying on explainability methods. This would enable producing human-readable interpretations for the actions selected by GNN models. Also, some works from the ML community propose to deal with uncertainty by predicting the posterior probability on model estimates (e.g., Bayesian NNs). Another possibility is to design testing procedures that systematically determine the safe operational ranges of GNN-based products before deployment (e.g., maximum network size, traffic loads). These tests could be done in controlled testbeds, and would allow issuing certifications defining bounded confidence levels within certain operational ranges.
\end{itemize}

\section*{\large{Conclusion}}
As a final thought, we have already witnessed the revolution of convolutional NNs applied to computer vision, or recurrent NNs applied to natural language processing. The main reason behind these success stories is that they started to use NN models specifically tailored to understand the underlying data structure in their respective domains. In this context, GNNs can be the perfect partner to materialize the revolution of deep learning in the field of communication networks, as data in networks is pervasively represented as graphs.

Moreover, from a practical standpoint, GNNs can accurately generalize to other networks unseen during training. This represents a crucial aspect for achieving commercial data-driven solutions, as it allows offline training in controlled testbeds and creating products that are directly ready for deployment in production networks. Nevertheless, in this article we have raised some relevant technological challenges and potential applications that remain to be explored before achieving widespread adoption of GNN-based solutions for communication networks.

\section*{\large{Acknowledgment}}
This publication is part of the Spanish I+D+i project TRAINER-A (ref. PID2020-118011GB-C21), funded by MCIN/AEI/10.13039/501100011033. This work is also partially funded by the Catalan Institution for Research and Advanced Studies (ICREA) and the Secretariat for Universities and Research of the Ministry of Business and Knowledge of the Government of Catalonia and the European Social Fund.

\bibliographystyle{IEEEtran} 
\bibliography{bibliography}

\section*{Biographies}

\vskip -3\baselineskip plus -1fil
\begin{IEEEbiographynophoto}{Jos\'e Su\'arez-Varela} is a postdoctoral researcher in the Barcelona Neural Networking center, Universitat Polit\`ecnica de Catalunya, Spain. His main research interests are in Machine Learning applied to communication networks. He is currently working on the application of Graph Neural Networks and Deep Reinforcement Learning to network control and management.
\end{IEEEbiographynophoto}
\vskip -2\baselineskip plus -1fil

\begin{IEEEbiographynophoto}{Paul Almasan} received his B.Sc. (2017) and M.Sc. (2019) in Computer Science from the Universitat Polit\`ecnica de Catalunya, Spain. He is currently pursuing his Ph.D. at the Barcelona Neural Networking Center (BNN-UPC). His research interests are focused on Graph Neural Networks and Deep Reinforcement Learning for networking.
\end{IEEEbiographynophoto}
\vskip -2\baselineskip plus -1fil

\begin{IEEEbiographynophoto}{Miquel Ferriol-Galm\'es} received his B.Sc. in computer science (2018) and M.Sc. in data science (2020) from the Universitat Politécnica de Catalunya. He is currently pursuing a Ph.D. at the Barcelona Neural Networking center (BNN-UPC). His main research interests are in the application of Graph Neural Networks to computer networks.
\end{IEEEbiographynophoto}
\vskip -2\baselineskip plus -1fil

\begin{IEEEbiographynophoto}{Krzysztof Rusek}
is an assistant professor at  AGH. His main research interests are performance evaluation of telecommunications systems, machine learning and data mining. Currently, he is working on the applications of Graph Neural Networks and probabilistic modeling for performance evaluation of communications systems.
\end{IEEEbiographynophoto}
\vskip -2\baselineskip plus -1fil

\begin{IEEEbiographynophoto}{Fabien Geyer} is currently with Technical University of Munich (TUM) and Airbus Central Research \& Technologies in Munich working on methods for network analytics, network performances and architectures. His research interests include novel methods for data-driven networking, formal methods for performance evaluation and modeling of networks.
\end{IEEEbiographynophoto}
\vskip -2\baselineskip plus -1fil

\begin{IEEEbiographynophoto}{Xiangle Cheng} received the Ph.D. degree in Computer Science at the University of Exeter, UK. He is currently a research fellow at Huawei. His research interests include 5G, Network AI, Stochastic \& Neural Combinatorial Optimization, Intelligent Wireless Networks and Mobile Computing, and Information Dynamics.
\end{IEEEbiographynophoto}
\vskip -2\baselineskip plus -1fil

\begin{IEEEbiographynophoto}{Xiang Shi} received her Bachelor's degree from the Minzu University of China, in 2014, and her PhD degree from the Institute of Computing Technology, Chinese Academy of Sciences in 2020. Currently she works in the Network Technology Laboratory at Huawei Technologies.
\end{IEEEbiographynophoto}
\vskip -2\baselineskip plus -1fil

\begin{IEEEbiographynophoto}{Shihan Xiao} received the Ph.D. degree in the Department of Computer Science and Technology, Tsinghua University, Beijing, China, in 2017. He is currently a technical expert of Network AI at Huawei Technologies. His research interests are in the areas of networking and machine learning.
\end{IEEEbiographynophoto}
\vskip -2\baselineskip plus -1fil

\begin{IEEEbiographynophoto}{Franco Scarselli} is an associate professor at the Department of Information Engineering and Mathematics, University of Siena, Italy. His research is in the field of machine learning with a particular focus on neural networks, machine learning for graphs and approximation theory. Applied research interests include also image understanding, information retrieval and bioinformatics.
\end{IEEEbiographynophoto}
\vskip -2\baselineskip plus -1fil

\begin{IEEEbiographynophoto}{Albert Cabellos-Aparicio} is a full professor at Universitat Politécnica de Catalunya, where he obtained his Ph.D. in computer science in 2008. He is director of the Barcelona Neural Networking center (BNN-UPC) and scientific director of the NaNoNetworking Center in Catalunya.
\end{IEEEbiographynophoto}
\vskip -2\baselineskip plus -1fil

\begin{IEEEbiographynophoto}{Pere Barlet-Ros}
is an associate professor at Universitat Politécnica de Catalunya and scientific director of the Barcelona Neural Networking center (BNN-UPC). From 2013 to 2018, he was co-founder and chairman of the machine learning startup Talaia Networks, which was acquired by Auvik Networks in 2018.
\end{IEEEbiographynophoto}
\vskip -2\baselineskip plus -1fil

\end{document}